\documentclass[showpacs,notitlepage,groupedaddress,superscriptaddress,twocolumn,numerical,nobalancelastpage,floatfix]{revtex4-2}
\usepackage[utf8x]{inputenc}
\usepackage{amsfonts,amsmath,amssymb,stmaryrd}
\usepackage{graphicx}
\usepackage{siunitx}
\usepackage[hidelinks]{hyperref}
\usepackage[nameinlink]{cleveref}
\Crefname{equation}{Eq.}{Eqs.}
\Crefname{figure}{Fig.}{Figs.}
\Crefname{tabular}{Tab.}{Tabs.}
\Crefname{table}{Tab.}{Tabs.}
\usepackage{nicefrac}
\usepackage{tikz}
\usepackage{tabularx}
\usepackage{bm} 
\usepackage{color}
\usepackage{scalerel}
\usepackage[normalem]{ulem}

\newcommand{\er}[1]{{\color{black}#1}}

\newcommand{\eq}{\begin{eqnarray}} 
\newcommand{\en}{\end{eqnarray}}

\DeclareSIUnit\gauss{G}
\DeclareSIUnit\bohrradii{a_0}

\begin{document}

\title{Making statistics work: a quantum engine in the BEC-BCS crossover}

\author{Jennifer~Koch}
\affiliation{Department of Physics and Research Center OPTIMAS, Technische Universit\"at Kaiserslautern, Germany}

\author{Keerthy~Menon}
\affiliation{OIST Graduate University, Onna, Okinawa, Japan}
	
\author{Eloisa~Cuestas}
\affiliation{Enrique Gaviola Institute of Physics, National Scientific and Technical Research Council of Argentina and National University of Córdoba, Córdoba, Argentina}
\affiliation{OIST Graduate University, Onna, Okinawa, Japan}

\author{Sian~Barbosa}
\affiliation{Department of Physics and Research Center OPTIMAS, Technische Universit\"at Kaiserslautern, Germany}
    
\author{Eric~Lutz}
\affiliation{Institute for Theoretical Physics I, University of Stuttgart, Stuttgart, Germany}
	
\author{Thom\'{a}s~Fogarty}
\affiliation{OIST Graduate University, Onna, Okinawa, Japan}

\author{Thomas~Busch}
\affiliation{OIST Graduate University, Onna, Okinawa, Japan}
	
\author{Artur~Widera}
\affiliation{Department of Physics and Research Center OPTIMAS, Technische Universit\"at Kaiserslautern, Germany}

\date{\today}

\begin{abstract}
Heat engines convert thermal energy into mechanical work both in the classical and quantum regimes. 
However, quantum theory offers  genuine nonclassical forms of energy, different from heat, which so far have not been exploited in cyclic engines to produce useful work.
We here experimentally realize a novel quantum many-body engine fuelled by the energy difference between fermionic and bosonic ensembles of ultracold  particles that follows from the Pauli exclusion principle. We employ a harmonically trapped superfluid gas of $^6$Li atoms close to a magnetic Feshbach resonance which allows us to effectively change the quantum statistics from Bose-Einstein to Fermi-Dirac.
We replace the traditional heating and cooling strokes of a quantum Otto cycle by  tuning the gas between a Bose-Einstein condensate of bosonic molecules and a unitary Fermi gas (and back) through a magnetic field. 
The quantum nature of such a Pauli engine is revealed by contrasting it to a classical thermal engine and to a purely interaction-driven device. We obtain a work output of several $10^6$ vibrational quanta per cycle with  an efficiency of up to 25\%. 
Our findings establish quantum statistics as a useful thermodynamic resource for work production, shifting the paradigm of energy-conversion devices to a new class of emergent quantum engines.
\end{abstract}

\maketitle

\begin{figure*}[t!]
    \includegraphics[scale=.45]{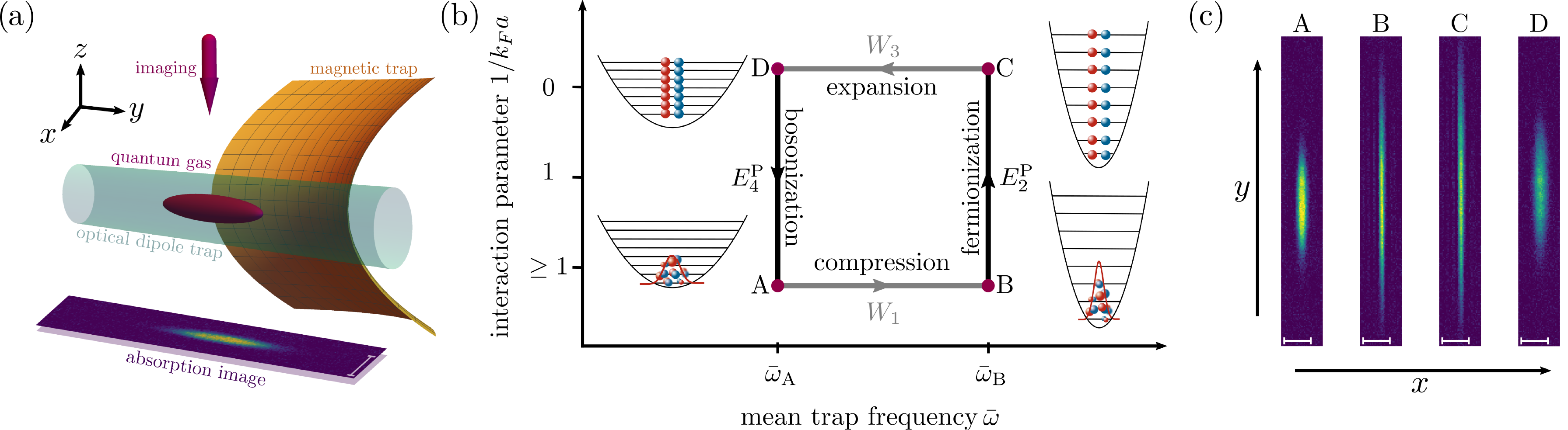}
    \caption{\textbf{Principles of the quantum Pauli engine.} (a) Schematic of the experimental setup. The atom cloud (purple ellipsoid) is trapped in the combined fields of a magnetic saddle potential (orange surface) and an optical dipole trap potential (blue cylinder) operating at a wavelength of \SI{1070}{nm}. The absorption pictures are taken with an imaging beam (purple arrow) in the $-z$-direction. The scale on the absorption picture corresponds to \SI{50}{\mu m}. (b)  Cycle of the quantum Pauli engine. Starting with an mBEC that macroscopically populated the ground state of the trap at well-defined temperature $T$ (point A), the first step, A$\rightarrow$B, performs work $W_1$ on the system by compressing the cloud through an increase of the radial trap frequency $\bar{\omega}_\text{B}>\bar{\omega}_\text{A}$. 
This is achieved by enhancing the power of the trapping laser. 
The second stroke, B$\rightarrow$C, increases the magnetic field strength from $B_\text{A} = 763.6 \, \text{G}$ to the resonant field $B_\text{C} = 832.2 \, \text{G}$, while keeping the trap frequency constant. 
This leads to a change in the quantum statistics of the system as the working medium now forms a Fermi sea with an associated addition of Pauli energy $E^\text{P}_2$, {which substitutes} the heat stroke. Step C$\rightarrow$D expands the trap back to the frequency  $\bar{\omega}_\text{A}$ and corresponds to the second work stroke  $W_3$. Finally, the system is brought back to the initial state with bosonic quantum statistics during step D$\rightarrow$A by driving the magnetic field from the resonant value $B_\text{C}$ to $B_\text{A}$, corresponding to a change in the Pauli energy $E^\text{P}_4$. The population distributions in the harmonic trap of the atoms with spin up (blue) and down (red) are indicated at each corner.  (c) Examples of absorption pictures at each point of the engine cycle,  where the particular change in size from B$\rightarrow$C is due to the Pauli stroke {indicating} that the Pauli energy increases the size of the cloud in the external potential.}  
    \label{fig_1_scheme}
\end{figure*}
%

Work and heat are two fundamental forms of energy transfer in thermodynamics. 
Work corresponds to the energy change induced by the variation of a mechanical parameter, such as the position of a piston, whereas heat is associated with the energy exchanged with a thermal bath \cite{cal85}. 
From a microscopic point of view,  work in quantum systems is related to a displacement of energy levels and heat to a modification of their population's probability distribution \cite{ali79}. 
Quantum heat engines realized so far convert thermal energy into mechanical work by cyclically operating between  effective thermal reservoirs at different temperatures \cite{zou17,kla19,ass19,pet19,bou21,kim22} like their classical counterparts, where heating/cooling strokes redistribute the quantum state populations. 
However, owing to the existence of distinct particle (Fermi or Bose) statistics, the level occupation probabilities of quantum many-body systems may strongly differ at the same temperature \cite{cal85}. 

At ultralow temperatures, in the quantum degenerate regime, an ensemble of indistinguishable bosonic particles will   all accumulate in the ground state, while fermionic systems will occupy quantum states with increasing energy due to the Pauli exclusion principle \cite{mas05}.  
This results in a profound energy difference due to the particles' quantum statistics, which is intimately linked to the symmetry of the many-body quantum wavefunction and originates from the spin-value of the particles.
The spin value, and hence the quantum statistics, is a fundamental property of a particle, and therefore not easily modified. 
However, a change in  quantum statistics can be  achieved in interacting atomic Fermi gases using Feshbach resonances \cite{RevModPhys.82.1225}. 
In such  systems, the crossover between a Bardeen-Cooper-Schrieffer (BCS) state of pairs of fermions to a molecular Bose-Einstein-condensate (mBEC) state of diatomic bosonic molecules can be experimentally realized by tuning an external magnetic field \cite{regal_2003_nature,Jochim_2003,Zwierlein_2003,regal_2004,Petrov_2004,bartenstein_2004,Zwerger_book}. 

Here we report the experimental realization of a novel many-body quantum engine ("Pauli engine") where the coupling to a hot/cold thermal bath -- and the resulting variation of temperature -- is replaced by a change of the quantum statistics of the system, from  Bose-Einstein to Fermi-Dirac  (and back). 
This engine cyclically converts energy stemming from the Pauli exclusion principle ("Pauli energy") into mechanical work. 
Its mechanism is of purely quantum origin, since the difference between fermions and bosons disappears in the classical high-temperature limit. 
We specifically employ an ultracold two-component Fermi gas of $^6$Li atoms confined in a combined optic-magnetic trap  \cite{met99} (Fig.~1),  prepared  close to a magnetic Feshbach resonance. 
We implement an Otto-like cycle \cite{kos17} by varying the trap frequency via the power of the laser forming the trapping potential, hence performing work. 
We further adiabatically change the magnetic field  through the crossover at constant trap frequency, thus changing the quantum statistics and the associated occupation probabilities. 
This step leads to the exchange of Pauli energy instead of heat. 
We emphasize that all strokes are unitary, preserving the entropy of the system throughout the cycle, in contrast to conventional quantum heat engines \cite{zou17,kla19,ass19,pet19,bou21,kim22}.  
We measure atom numbers and cloud radii from \textit{in-situ} absorption images \cite{foot2005}, from which we determine the energy of the gas after each stroke of the engine. 
We employ the latter quantities to evaluate both efficiency and  work output of the Pauli engine, analyze its thermodynamic performance when parameters are varied, and compare it to theoretical calculations. 

It is instructive to begin with a discussion of the simple case of a 1D harmonically trapped noninteracting ideal gas at zero temperature in order to gain physical insight into the energetic potential of the Pauli  principle \cite{mas05}. 
A fully bosonic system only populates the ground state with energy $E^\mathrm{B}=N \hbar {\omega}/2$, where $N$ is the number of particles and $\omega$ is the trap frequency \cite{cal85}. 
By contrast, the fermionic counterpart populates all the energy levels up to the Fermi energy $E_F= \hbar {\omega}(2N-1)/2$ and the total energy of the system is accordingly $E^\mathrm{F}={\hbar {\omega}} (N +1/2)^2/2$ \cite{cal85}. 
The resulting energy difference due to the change of quantum statistics, the Pauli energy, is therefore $E^\text{P}= E^\mathrm{F}- E^\mathrm{B}={\hbar {\omega}} (N^2+N+1)/2$. This enormous difference of total energy at zero temperature originates from the underlying quantum statistics, dictating a population probability distribution across the available quantum energy levels $E_n$ according to
$f_n =1/[\exp[({E_{n}-\mu})/({k_B T)]}\pm 1]$, 
where the $+$ ($-$) sign in the denominator is for fermions (bosons) with half-integer (integer) spin \cite{pauli_1940}; here $\mu$ is the chemical potential and $k_B$ is the Boltzmann constant.
 For increasing temperature, both these distribution functions reduce to the well-known Boltzmann factor, and quantum effects are absent.
Microscopically, the Pauli energy is equal to $E^\text{P}=\sum_n \Delta f_n E_n$, an expression reminiscent of that of heat for systems coupled to a bath \cite{ali79}. Importantly, the quadratic dependence of the energy difference on the particle number between BEC and Fermi sea  in the 1D case implies that the Pauli energy can be significant for large $N$,  far exceeding  typical energy scales in comparable quantum thermal machines.

\begin{figure*}[t]
    \centering
    \includegraphics[scale=.5]{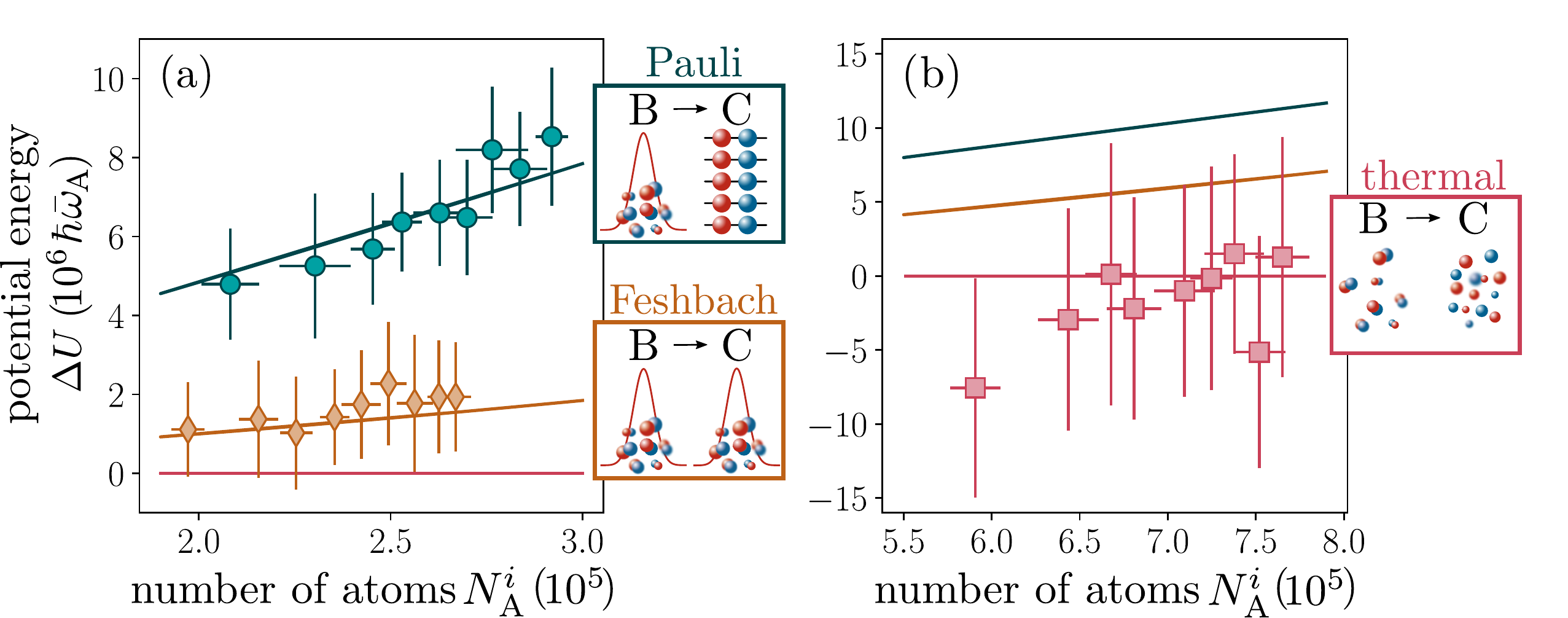}
    \caption{\textbf{Contribution of the Pauli energy.} Mean potential energy variation $\Delta U$ as a function of the number of atoms in a single spin state $N^i_{\mathrm{A}}$ for the magnetic field change between cycle points $\mathrm{B}\rightarrow\mathrm{C}$ for a gas performing a {Pauli stroke (bringing the gas from an mBEC to a unitary gas) (cyan) a Feshbach stroke (always remaining in the mBEC regime) (orange), and a thermal stroke for the same magnetic fields as in the Pauli stroke (red)}. Data points represent experimental measurements, solid lines are predictions of our model.
    Due to the much increased temperature of the gas in the thermal case, the trap depth and compression ratio have been chosen differently in the experimental realization to be (a) for the Pauli and Feshbach strokes $T=120\,$nK, $T/T_F=0.3$ (measured in the mBEC regime), and $\bar{\omega}_\mathrm{B}/\bar{\omega}_\mathrm{A}=1.5$; and (b) for the thermal stroke $T \approx \SI{1150}{nK}$, $T/T_F \approx 0.7$ (measured in the mBEC regime), and $\bar{\omega}_\mathrm{B}/\bar{\omega}_\mathrm{A}=1.1$.
    The insets show microscopic sketches of the quantum state of the gas, transitioning between mBEC and Fermi sea, remaining in the mBEC, or transitioning between a gas of free molecules to a gas of free atoms, respectively. The error bars for all data points denote the standard deviation of 20 repetitions.
    }
    \label{fig_2_pauli_stroke}
\end{figure*}

\subsection*{Controlling the quantum statistics}

We prepare an interacting, three-dimensional quantum-degenerate two-component Fermi gas of up to $N=6 \times 10^5$ $^6$Li atoms (Fig.~\ref{fig_1_scheme}(a)) \cite{doi:10.1063/1.5045827}, with equal population $N^i$ of two lowest-lying Zeeman states $i$, i.e., $N^i=N/2$. 
A broad Feshbach resonance centered at $832.2\, \text{G}$ \cite{PhysRevLett.110.135301} allows us to change the nature of the many-body state: an mBEC of $N/2$ molecules is formed at magnetic field strengths below the resonance, whereas on resonance a strongly interacting Fermi sea of $N$ $^6$Li atoms emerges.
For the bosonic regime, we operate at  $B_{\mathrm{A}}=763.6\,\text{G}=B_{\mathrm{B}}$, where the mBEC has an interaction parameter of $1/k_F a \approx 2.3$, with $k_F$ being the Fermi wavevector and $a$ the $s$-wave scattering length \cite{dalfovo_review_1999,BEC_stringari}. 
Here, the temperature of the gas is about  $T=120\, \text{nK}$, corresponding to $T/T_F \approx 0.3$ with the Fermi temperature $T_F  = \hbar \bar{\omega}(3N)^{1/3}/k_B$, and $\bar{\omega}$
the geometric mean trap frequency which can be  experimentally controlled.
The unitary regime appears on resonance, $B_{\mathrm{C}}=832.2\,\text{G}=B_\mathrm{D}$,
where the gas is dilute, but strongly interacting, and exhibits universal behavior, which is independent of the microscopic details due to the divergence of the scattering length,  $1/k_F a =0$ \cite{giorgini_review_2008,Zwerger_book}. 
In this regime, Pauli blocking of occupied single-particle states leads to Fermi-Dirac-type statistics \cite{enrico_fermi_school_CXCI,bouvrie_2017,bouvrie_2019,yongshi_1994}. 
When the mBEC is adiabatically ramped to unitarity, the reduced temperature drops to values well below $T/T_F < 0.2$ \cite{chen_thermo_2005}.
For magnetic field values above the Feshbach resonance, the gas enters the regime of a BCS superfluid, which is not considered in this work.

The experimental implementation of the quantum Otto cycle ABCD \cite{kos17}  starts with a mBEC and employs alternating strokes changing the mean trap frequency via the dipole trap laser intensity, and the quantum statistics via the external magnetic field (Fig.~\ref{fig_1_scheme}(b)).  
The  novel Pauli stroke B$\rightarrow$C, associated with the Pauli energy change $E^P_2$, replaces the traditional heating by a bath with the change of quantum statistics.
All the strokes are unitary (apart from minor particle losses, see below) and,  therefore,   the entropy of the system is constant throughout the cycle, in contrast to conventional quantum heat engines \cite{zou17,kla19,ass19,pet19,bou21,kim22}. 
The quantum gas can be analyzed at every instance along the cycle by  taking \textit{in-situ} absorption images in the  optic-magnetic trap \cite{foot2005}. 

\begin{figure*}[tb]
    \centering
    \includegraphics[scale=.45]{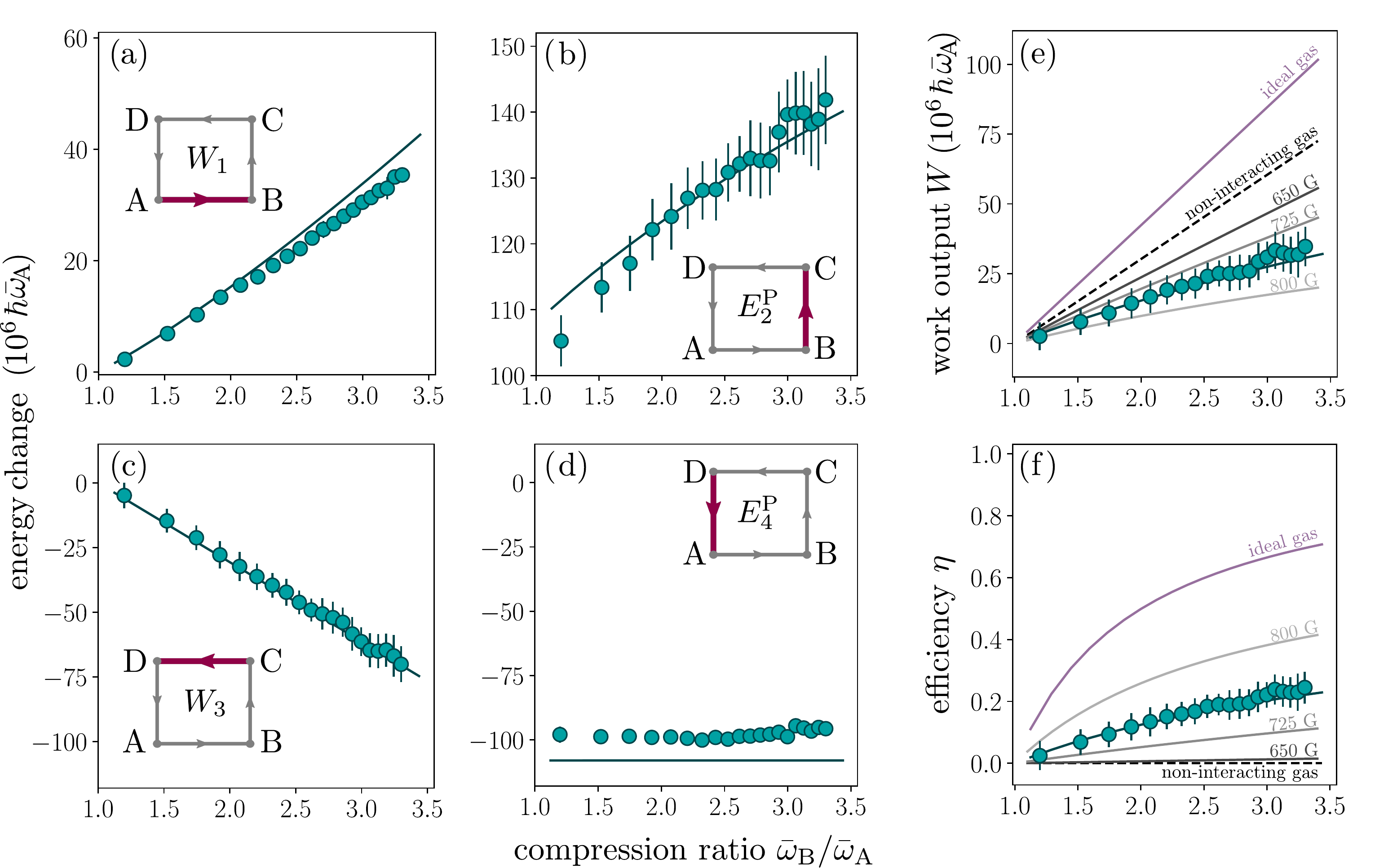}
    \caption{\textbf{Performance of the Pauli engine.} (a)-(d) Work outputs, $W_1$ and $W_3$, and Pauli energies, $E^\text{P}_2$ and $E^\text{P}_4$, as a function of the compression ratio $\bar{\omega}_\mathrm{B} / \, \bar{\omega}_\mathrm{A}$  for fixed $\bar{\omega}_\mathrm{A}$ and fixed number of atoms $N^i_{\text{A}}\approx2.5 \times 10^5$ at point A. The experimental data points (cyan) are the mean value of 20 repetitions and the error bars indicate their statistical uncertainty. The numerical calculations are indicated by the (cyan) solid lines. The insets show the corresponding stroke. (e)-(f) Work output $W$ and efficiency $\eta$. The experimental points and the numerical simulations are in cyan. For comparison, the $W$ and $\eta$ values for a noninteracting Li-gas are depicted as black dashed lines. They have been obtained by setting the $s$-wave scattering length $a$ to zero for points A and B (this limit corresponds to a magnetic field $B_{\text{A}}$ far below the resonance leading to point-like composite bosons with infinite binding energy) and using the unitarity formulas for points C and D. Also, $W$ and $\eta$ for an ideal gas are shown as purple solid lines and mark the upper bounds of the machine. They have been obtained by using the energy of ideal gases in the degenerate regime (Bose gas for points A and B and Fermi gas for points C and D). The gray solid lines are the theoretical values obtained for different magnetic fields: $B_{\text{A}} = 800,\, 725,\text{ and } 650  \text{ G}$ from lighter to darker (dimer-dimer interaction strength $g/g_0=2.53,\, 0.51, \text{ and } 0.16$, with $g_0$ being the interaction strength for $B_{\text{A}}=763.3\text{ G}$).}
    \label{fig_3_pauli_engine_omega}
\end{figure*}


\subsection*{Quantum contributions to the engine operation}

Let us first analyze the Pauli stroke B$\rightarrow$C of the engine cycle during which the change of quantum statistics takes place.
Contrary to the ideal gas case discussed above,  where a pure change of the quantum statistics was considered, the atoms in the experimental system are not at zero temperature and further interact with a strength that depends on the magnitude of the magnetic field.
Both of these have an effect on the cloud sizes measured. To isolate the different contributions, and to illustrate the quantum contribution of this stroke, 
we evaluate the energy increase indicated by the cloud size in the trapping potential for different scenarios. 
In order to obtain a model-independent evaluation, we extract the increased potential energy from experimentally obtained \textit{in-situ} absorption images. Exploiting the radial symmetry of the cloud with radial trap frequencies $\omega_x \approx \omega_z$ and the known shape of the optical potential, we compute the mean potential energy of the gas in the trap, for a given atom number $N$, as 
 $ U  = ( m \omega^2_x \langle x^2 \rangle +  m \omega^2_y/2 \langle y^2 \rangle) N$ \cite{PhysRevA.78.013630}; here,  $m$ is the mass of a $^6$Li atom and $\langle x^2 \rangle$, $\langle y^2 \rangle$ are the measured mean-square sizes of the cloud in the $x$, $y$ directions \cite{Thomas_2005, PhysRevA.78.013630,TAN20082987,Zwerger_book}. 
Taking the difference of energies before and after the stroke, we compute the stroke-induced increase in {potential} energy $\Delta U $ of the gas (Fig.~\ref{fig_2_pauli_stroke}).
In Fig.~\ref{fig_2_pauli_stroke}(a), we compare this energy change for a Pauli stroke, simultaneously varying quantum statistics and interaction, with a so-called Feshbach stroke \cite{Li:18,Keller:20}, where the interaction of the gas is modified while preserving the {effective} quantum statistics by always remaining in the mBEC state. 
Moreover, we contrast the Pauli stroke behavior to a gas at a much higher temperature of $T/T_F \approx 0.7$, repeating the magnetic field ramp of the Pauli stroke, and hence the corresponding scattering length, but where  quantum effects {are} suppressed (Fig.~\ref{fig_2_pauli_stroke}(b)). 
We observe that the change of quantum statistics during the Pauli stroke yields a much larger difference of potential energy and a much faster increase with the particle number. The variation of potential energy is hence mostly due to the Pauli energy. 
We also note good agreement with the theoretical curves (solid lines, see Methods) obtained using analytical formulas in the Thomas-Fermi regime for the mBEC \cite{dalfovo_review_1999}, and the known energy expressions for a strongly interacting Fermi gas at unitarity \cite{giorgini_review_2008}. 
The scaling of the Pauli energy with particle number deviates from the simple example of a 1D ideal gas, as discussed  below. 

%
\subsection*{Performance of the quantum many-body engine}
To experimentally evaluate the performance of the  Pauli engine, we extract the energy change for every cycle stroke from absorption images as before (Fig.~\ref{fig_3_pauli_engine_omega}).
From these energy contributions,  the performance of the engine can then be characterized by computing the work output $W$ and efficiency $\eta$ \cite{cal85}, which are defined in analogy to the standard quantum Otto cycle as \cite{kos17}
%
\begin{equation}
W= -\left( W_1 +  W_3\right) \quad \text{and} \quad  \eta = {W}/{E^\text{P}_2}.
\end{equation}
%
\begin{figure*}[t!]
    \centering
    \includegraphics[scale=.45]{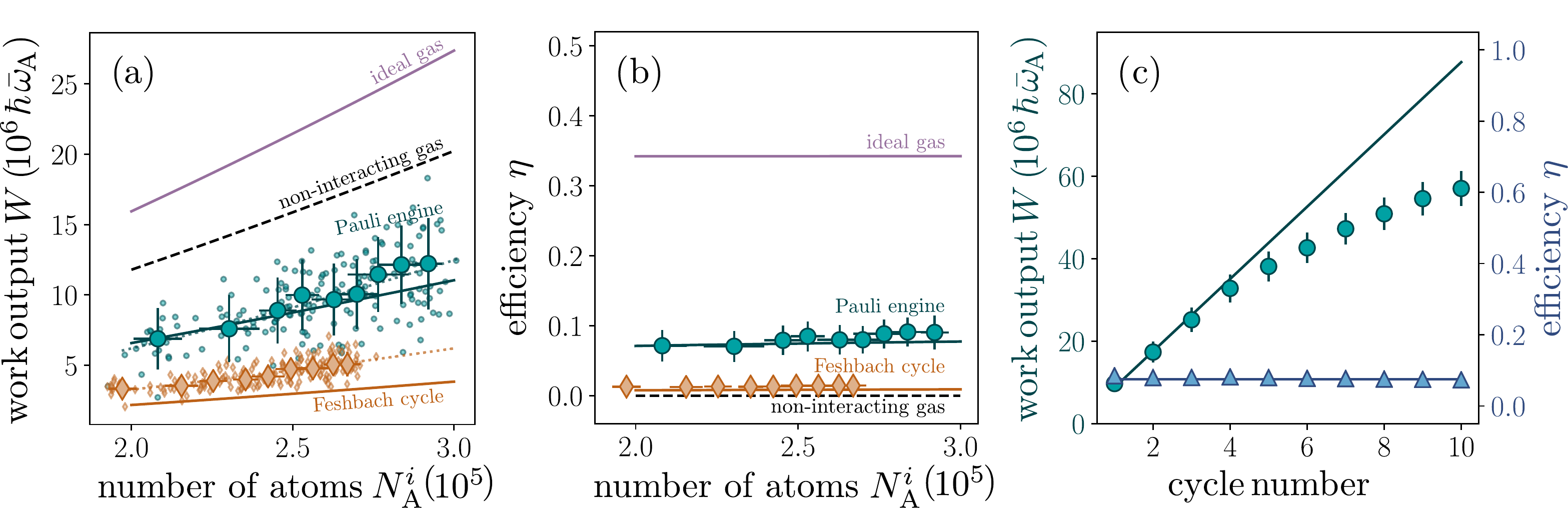}
    \caption{\textbf{Comparison of the performance of quantum many-body devices.} (a){-}(b) {W}ork  output and efficiency as a function of the number of atoms $N^i_{\mathrm{A}}$ in one spin state for a compression ratio $\bar{\omega}_\mathrm{B} / \, \bar{\omega}_\mathrm{A} = 1.5$ for the Pauli engine (cyan) and the Feshbach cycle (orange). 
    Small symbols denote individual realizations, large symbols indicate averages of 20 repetitions with error bars spanning one standard deviation.
    For comparison, calculations for a non-interacting gas (black dashed line) and an ideal gas (purple solid line) are shown. 
    Dotted lines are fits to the respective work output $W\propto N^\alpha$.
      The {fitted} exponents $\alpha$ are 1.73(18) and 1.58(13) for the Pauli engine and Feshbach cycle, respectively, well reproduced by our model. 
    Importantly, the efficiency of the Feshbach cycle is essentially zero. 
    (c) Accumulated work output (cyan) and efficiency (blue) of the Pauli engine over several cycles for a compression ratio $\bar{\omega}_\mathrm{B} / \, \bar{\omega}_\mathrm{A} = 1.5$ and an initial number of atoms in one spin state of about $N^i=2.5 \times 10^5$. Also here, symbols are the mean of 20 repetitions, error bars denote one standard deviation.}
    \label{fig_4_pauli_engine_N}
\end{figure*}
%
The energy changes for individual strokes (Figs.~\ref{fig_3_pauli_engine_omega}(a)-(d)) allow one to obtain an intuitive understanding of the performance. The two work strokes (a) and (c) simply follow the variation  of  the trap frequency. The important Pauli-stroke energy change (b) directly reflects the growing Fermi energy in steeper potentials for constant particle numbers, pointing toward increased work output of an engine for increased compression ratio. {The complementary stroke (d) is independent of this change because the trap frequency $\bar \omega_A$ is fixed in the experiment.}

We  observe that the work output increases with compression ratio (\er{Fig.}~\ref{fig_3_pauli_engine_omega} (e)), paralleled by an increased efficiency $\eta$ for the compression ratios experimentally accessible.
The corresponding performance of the engine for increasing particle number, but fixed compression ratio, is shown in Fig.~\ref{fig_4_pauli_engine_N}(a)-(c) together with the work output for an increasing number of cycles. 
We find that our realization of the Pauli engine produces a total  work output of several $10^6\,\hbar \bar{\omega}_\mathrm{A}$ and significant efficiencies above $10\%$ for compression ratios larger than $1.5$. 
The largest efficiency achieved in the experiment is 25$\%$, while for a very large compression ratio of about $10$ (not accessible experimentally), the theoretical efficiency for the same experimental parameters can be higher than 50\%. 
Moreover, we note that $W$ increases with $N$, whereas $\eta$ remains constant at about \SI{7}{\%} for the chosen parameters. 
Also, the comparison between a quantum-statistics driven Pauli engine  and an interaction-driven Feshbach device shows superior performance of the  Pauli engine in Fig.~\ref{fig_4_pauli_engine_N}, where specifically the efficiency is essentially zero for the Feshbach cycle.
Finally, when operating the engine over multiple subsequent cycles, the efficiency does not change, while the work output increases after each cycle almost linearly, until atom losses reduce the total work output for more than five cycles (Fig.~\ref{fig_4_pauli_engine_N}(c)). 

\subsection*{Theoretical analysis}

Additional insight about the performance and limitations of the Pauli engine can be gained from a model of the engine performance.
Here, all energy contributions and the fact that the change of quantum statistics is inevitably connected to a change of interaction strength have to be included. 
The molecules interact with each other via a contact interaction \cite{dalfovo_review_1999}, 
which can be quantified using the interaction strength $g=1.2 \pi\hbar^2 a/m$ \cite{Petrov_2004}, with $a$ being the $s$-wave scattering length.
The ground-state energy of the bosonic system is then found by numerically solving the Gross-Pitaevskii equation, which can be used to study any condensate with sufficiently weak interaction. 
Additionally, for our experimental parameters of the mBEC, the Thomas-Fermi approximation holds, and therefore we also rely on analytical formulas for the bosonic energy (see Methods and \cite{dalfovo_review_1999}). 
Once we tune the magnetic field to resonance, the system evolves into a strongly interacting unitary Fermi gas, whose energy is given by $E_{\mathrm{res}}= 3 \sqrt{1+\beta} N E_F/4$, with $\beta = -0.49$ and the Fermi energy $E_F = k_B T_F$ (see Methods, and \cite{giorgini_review_2008, kinast_science_2005, PhysRevLett.110.135301}). 
Since the energy at unitarity is almost independent of temperature for $T/T_F<0.2$ (which is satisfied in our experiments), this zero-temperature expression provides a good approximation. 
Furthermore, it is important to consider the molecular binding energy $E_{\mathrm{m}} = - \hbar^2/(m a^2)$ associated with dissociating a molecule into two atoms. 
While it has no consequences for the work output, because the $s$-wave scattering length does not change during the work strokes, it still has to be provided to the system during the Pauli stroke. 
Hence, it must be included in the energy-cost calculation of the engine's efficiency. 
These calculations yield the theory curves in Figs.~\ref{fig_3_pauli_engine_omega} and \ref{fig_4_pauli_engine_N}.


Beyond the very good agreement with our data the {theoretical calculations} additionally provide limits on the engine performance. 
In particular, the simple ideal-gas model leads to upper bounds for work output and efficiency given by
$W_\text{ideal} = (3 N)^{4/3} \hbar (\bar{\omega}{_\text{B}}-\bar{\omega}{_\text{A}})/4$ and $\eta_\text{ideal} = 1 -\bar{\omega}{_\text{A}}/\bar{\omega}_{\text{B}}$. 
Here, the exponent of the work-output with particle number is different from the one-dimensional case due to a modified density of states, i.e., number of available single-particle states in the potential. 
The ideal efficiency is similar to the maximum Otto efficiency \cite{kos17}. 
Compared to these upper limits (Figs.~\ref{fig_3_pauli_engine_omega} and \ref{fig_4_pauli_engine_N}), we find that the experimental system shows reduced values but of the same order of magnitude.
Moreover, our model allows to predict the performance for different magnetic field values in the mBEC regime (Fig.~\ref{fig_3_pauli_engine_omega}(e), (f)),  where the work output is reduced with increased initial {effective} repulsion of the mBEC, while the efficiency increases. 
This scaling of the work output is a consequence of the competition between  interactions between molecules in the initial mBEC state and the effect of changing the quantum statistics. 
For stronger initial {effective} repulsion, the mBEC cloud already exhibits a relatively large energy in the trap, so that the change of quantum statistics can only contribute a comparatively smaller amount of energy during the Pauli stroke. This suggests an optimal work output for an initially noninteracting mBEC.
By contrast, for the efficiency, the binding energy has to be provided to the system when dissociating a molecule into two atoms. This binding energy quickly grows as the magnetic field deviates from the resonance value, and the associated energy cost quickly reduces the efficiency. For the experimentally inaccessible case of a noninteracting mBEC, this binding energy is so large that the efficiency of the Pauli energy is essentially zero. 
These considerations point to an optimal point of operation which might additionally be temperature dependent. 

An important result demonstrating the nonclassical drive of the Pauli engine is reproduced by both experimental findings and the theoretical model: the work output as a function of the number of particles $W \propto N^\alpha$ scales with a fitted exponent $\alpha=1.73(18)$ which is close to the prediction of 1.4 given by theory (Fig.~\ref{fig_4_pauli_engine_N}(a)). 
This exponent differs from the simple example of a 1D ideal gas in the introduction mainly due to a different density of states. 
While a similar exponent {is expected} for the Feshbach-driven stroke which remains always in the mBEC regime, the efficiency for this engine is close to zero.
Most importantly, the exponent is larger than one, which is the expected exponent for a noninteracting, classical gas \cite{cal85, williams_njp_2004, Beau16}.

\subsection*{Perspectives}

We have constructed an engine purely fuelled by quantum statistical effects induced by a Feshbach sweep that changes the nature of the working medium from a bosonic mBEC to a {strongly interacting} Fermi gas at unitarity. By replacing the traditional heating and cooling strokes of an Otto engine by a change of the symmetry of the wavefunction this quantum many-body engine represents a paradigm shift for quantum machines. The energy-state population of the working medium of a quantum engine may indeed not only be varied by the temperature of an external heat bath but also by a change of quantum statistics during the cycle. 
The Pauli energy associated with such a modification of quantum statistics may be large: in solids, the energy of electrons in the conduction band corresponds to thousands of Kelvin, much above usually accessible thermal energies. 
At the same time, since  the operation of the Pauli engine is in principle unitary, it is free from the usual  dissipation mechanisms of conventional quantum motors. A {relevant} question is hence to investigate how coupling this device to other quantum systems for work extraction might or might not induce dissipation.
Moreover, the three-dimensional working fluid shows a less favorable scaling with atom number  compared to the simple example of a 1D harmonic oscillator potential. In the future, it will thus be interesting to understand and optimize the engine's output, for instance, with respect to dimensionality, interaction strength and temperature. The Pauli engine furthermore provides a unique testbed to experimentally study the role of diabaticity and nonequilibrium dynamics on the performance of a quantum many-body machine \cite{Jaramillo_2016,Hartmann_2020,Fogarty:20}.


\section*{Acknowledgments}

This work was supported by the Deutsche Forschungsgemeinschaft (DFG, German Research Foundation) via the Collaborative Research Center Sonderforschungsbereich SFB/TR185 (Project 277625399) and Forschergruppe FOR 2724. It was also supported by the Okinawa Institute of Science and Technology Graduate University. J.K. was supported by the Max Planck Graduate Center with the Johannes Gutenberg-Universität Mainz. E.C. was supported by the Argentinian National Scientific and Technical Research Council (CONICET). TF was supported by JSPS KAKENHI-21K13856. We thank B.~Nagler and A.~Guthmann for carefully reading the manuscript.

\appendix
\section*{Appendix}

\renewcommand{\theequation}{M.\arabic{equation}}
\renewcommand{\thefigure}{M.\arabic{figure}}

\setcounter{figure}{0}

\paragraph*{Setup and sequence.}

We prepare a degenerate two-component fermionic $^6$Li gas in the two lowest-lying Zeeman substates of the electronic ground state $²S_{1/2}$ confined in an elongated quasi-harmonic trap comprising a magnetic and an optical dipole trap (ODT) operating at a wavelength of \SI{1070}{nm},  for details of the experimental setup see Ref.~\cite{doi:10.1063/1.5045827}.  
We cool the gas using evaporative cooling until reaching a temperature of \SI{120}{nK} in a trap with geometric mean trap frequency $\bar{\omega} = (\omega_x \omega_y \omega_z)^{1/3}$, where typical values used for the engine performance are tabled in Tab.~\ref{table_exp_parameters}.
Evaporation takes place on the BEC side of the crossover at a magnetic field of \SI{763.6}{G}.  
After evaporation, we hold the cloud for \SI{150}{ms} to let it equilibrate. 
By varying the loading time of the magneto-optical trap prior to evaporation, we adjust the number of atoms per spin state $N^i$ of the cloud between $1.75 \times 10^5$ up to $3.0 \times 10^5$.  

\begin{figure}[tb]
    \includegraphics[scale=1.1]{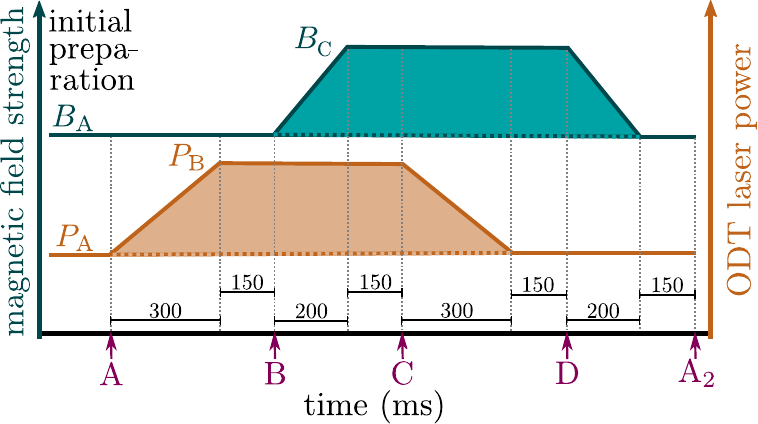}
    \caption{Sequence of the cycle. Magnetic field strength (cyan) and power of the ODT laser (orange). The imaging points during the cycle are marked with purple arrows.}  
    \label{fig:sequence}
\end{figure}

The quantum Otto cycle is implemented by alternating changes of the trap frequency through the dipole trap laser and of the interaction strength through the magnetic field close to the Feshbach resonance.
Figure \ref{fig:sequence} illustrates the experimental sequence of a single cycle. 
After initial preparation of the mBEC, we compress the trap adiabatically increasing the laser power of the ODT during \SI{300}{ms} from $P_{\mathrm{A}}$ to $P_{\mathrm{B}}$. 
The trap frequencies in both radial directions increase with the square root of the laser power of the ODT while the trap frequency along the axial direction remains insignificant compared to the magnetic-field's trapping frequency for all the powers used in this work. 
The resulting geometric mean trap frequencies are given in table \ref{table_exp_parameters}.
After a waiting time of \SI{150}{ms} with constant trap frequency, we reach cycle point B. 
Then, we adiabatically increase the magnetic field strength linearly from $B_{\mathrm{A}}$ to $B_{\mathrm{C}}$ to change the interaction strength during the stroke B$\rightarrow$C. 
Changing the magnetic field strength does not significantly alter the trap frequency in the axial direction. 
After a waiting time of \SI{150}{ms} with constant magnetic field, we reach point C. 
In the next step, we ramp the ODT-laser power to the initial value $P_{\mathrm{A}}$. Therefore, we expand the gas (C$\rightarrow$D) in the trap. 
Finally, we close the cycle with a second magnetic field ramp back to $B_{\mathrm{A}}$. 
After running an entire cycle, we verify that the measurement point $\mathrm{A}_2$ is equivalent to point A. 
In all the measurement series, we have $B_{\mathrm{A}}<B_{\mathrm{C}}$, $P_{\mathrm{A}}<P_{\mathrm{B}}$ and therefore, $\bar{\omega}_{\mathrm{A}}<\bar{\omega}_{\mathrm{B}}$ . Table~\ref{table_exp_parameters} depicts the experimental parameters for the three types of cycles we run. 
%
%
%
\begin{table}[htb]
\caption{Experimental settings of the three types of cycles for which we vary the number of atoms. For the case in which we change the compression ratio $\bar{\omega}_{\mathrm{B}}$ of the Pauli engine, the power of the ODT laser $P_{\mathrm{B}}$ is iterated from \SI{50}{mW} up to \SI{1}{W} in steps of \SI{50}{mW}. We extract the values of the $s$-wave scattering length $a$ from Ref.~\cite{PhysRevLett.110.135301}.}
    \begin{tabular}{c@{\hspace{0.5 cm}}c@{\hspace{0.3 cm}}c@{\hspace{0.3 cm}}c}
    \hline
    & Pauli &  mBEC-mBEC   & thermal \\ 
    \toprule              
    $B_{\mathrm{A}} \, (1/k_Fa)$  & \SI{763.6}{G} (2.3) & \SI{720}{G} (4.7) &    \SI{763.6}{G} (2.3)  \\ 
    $B_{\mathrm{C}} \, (1/k_Fa)$  & \SI{832.2}{G} (0.0) & \SI{780}{G} (1.6)  &     \SI{832.2}{G} (0.0) \\ 
    $P_{\mathrm{A}}$ & \SI{30}{mW} &  \SI{30}{mW}  & \SI{400}{mW} \\ 
    $P_{\mathrm{B}}$ & \SI{100}{mW} &  \SI{100}{mW} & \SI{500}{mW} \\ 
    $\bar{\omega}_{\mathrm{A}}/(2\pi)$ & \SI{84.97}{Hz} &  \SI{84.16}{Hz} & \SI{206.39}{Hz}\\ 
    $\bar{\omega}_{\mathrm{B}}/(2\pi)$ & \SI{129.29}{Hz} & \SI{128.06}{Hz} &  \SI{222.40}{Hz}\\ \hline
    \end{tabular}
    \label{table_exp_parameters}
\end{table}
%


\paragraph*{Atom-number measurement.}

To determine the number of atoms from \textit{in-situ} absorption pictures, we image the energetically higher-lying spin state $m_I = 1$ by standard absorption imaging on a CCD camera. This procedure is identical for all the regimes considered. 
Since the number of atoms of the spin state $m_I = 0$ is the same as for $m_I =1$, the total number of atoms $N$ (sum of atoms with spin up and down) is twice the number of atoms of the measured spin state $N^i$. 
Importantly in the case of an mBEC, the number of molecules is half the number of total atoms. 
We extract the  column density from the measured absorption pictures.  Adding up the density pixelwise and including the camera's pixel area and the imaging system's magnification, we obtain the number of atoms. We include additional laser-intensity-dependent corrections \cite{phdthesis_nagler}. The imaging system has a resolution of \SI{2.2}{\micro m}.

\begin{figure}[tb]
    \includegraphics[scale=0.7]{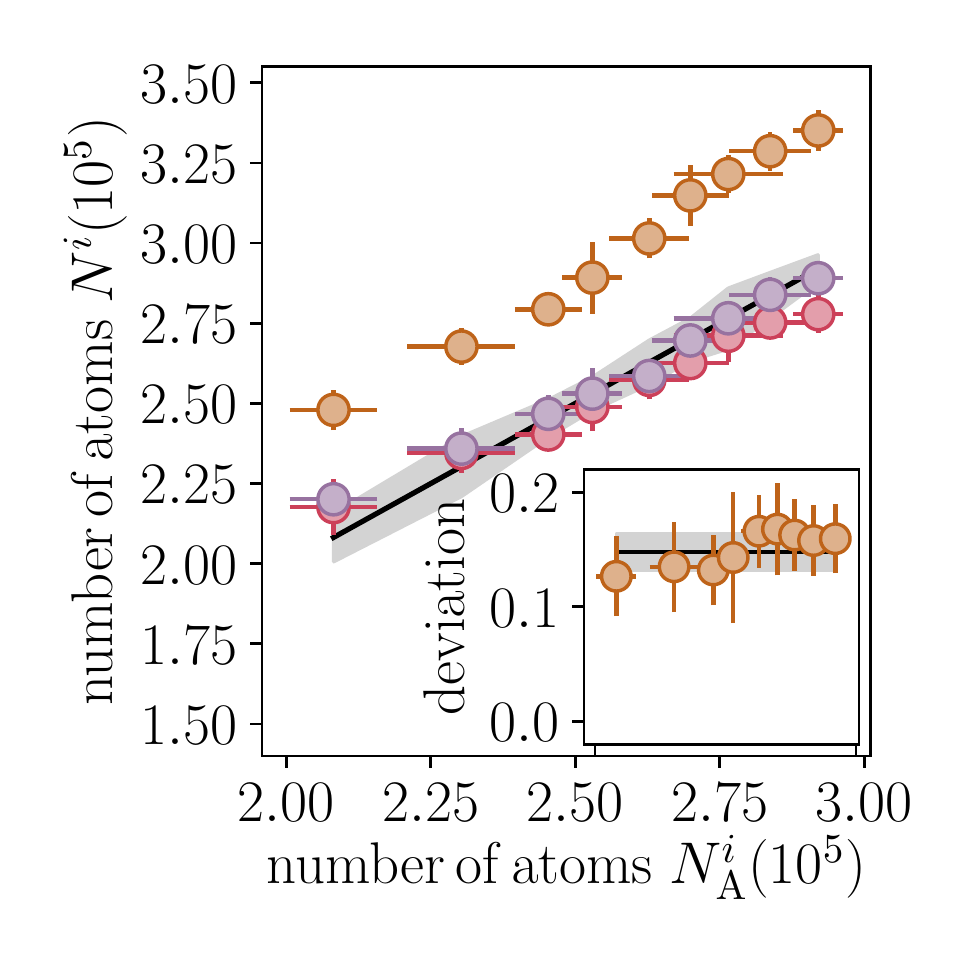}
    \caption{Determination of the correction factor of the number of atoms on resonance for the Pauli engine. Measured number of atoms of one spin state $N^i$ in point B (red), C (orange), and return to point $\mathrm{B_{return}}$ after reaching point C (purple) in dependence of the number of atoms of one spin state $N^i_{\mathrm{A}}$ in point A. The measured number of atoms in point C cannot differ from the number of atoms in points B and  $\mathrm{B_{return}}$, when the later are equal. Therefore, the measured number of atoms on resonance is higher than the actual number of atoms. Data points are averages of ten repetitions, uncertainties indicated one standard deviation. The inset shows the calculated deviation $N_{\mathrm{C}}/N_{\mathrm{B_{return}}} - 1$ (orange points) for the number of atoms on resonance in dependence of $N^i_{\mathrm{A}}$. The mean value of these deviations is \SI{15}{\percent} and is visible as a black line, their standard deviation is the grey area. The $x$-axis of the inset spans the same range as the main axis.   
    \label{fig:atom_number_correction_factor}}
\end{figure}

The optical absorption cross section changes its value throughout the BEC-BCS crossover, because it depends on the optical transition frequency \cite{foot2005} which is, in turn, a function of the magnetic field strength.
We determine the absorption cross section for a magnetic field of \SI{763.6}{G}, and we use this value for further magnetic field strengths. This leads to the fact that the measured number of atoms on resonance is higher than the actual number of atoms. Therefore, we determine a correction factor to compensate this imperfection in independent measurements at different magnetic field strengths, see Fig.~\ref{fig:atom_number_correction_factor}. 
We prepare an atomic cloud of varying atom number at $B_{\mathrm{A}} = \SI{763.6}{G}$. We then measure the atom number for this field at points A and B. We repeat the measurement for $B_{\mathrm{C}}$, and, to exclude atom loss we ramp the field from $B_{\mathrm{B}}=B_{\mathrm{A}}$ adiabatically to $B_{\mathrm{C}}$ and back to $B_{\mathrm{B}}$ before measure.
The atom numbers for $B_{\mathrm{B}}$ with and without the additional ramp to $\mathrm{C}$ are almost identical, from which we conclude that the number measured at $B_{\mathrm{C}}$ has to be the same. The correction factor is then calculated from the difference in atom number between these two data sets. 

\begin{figure}[tb]
    \includegraphics[scale=0.7]{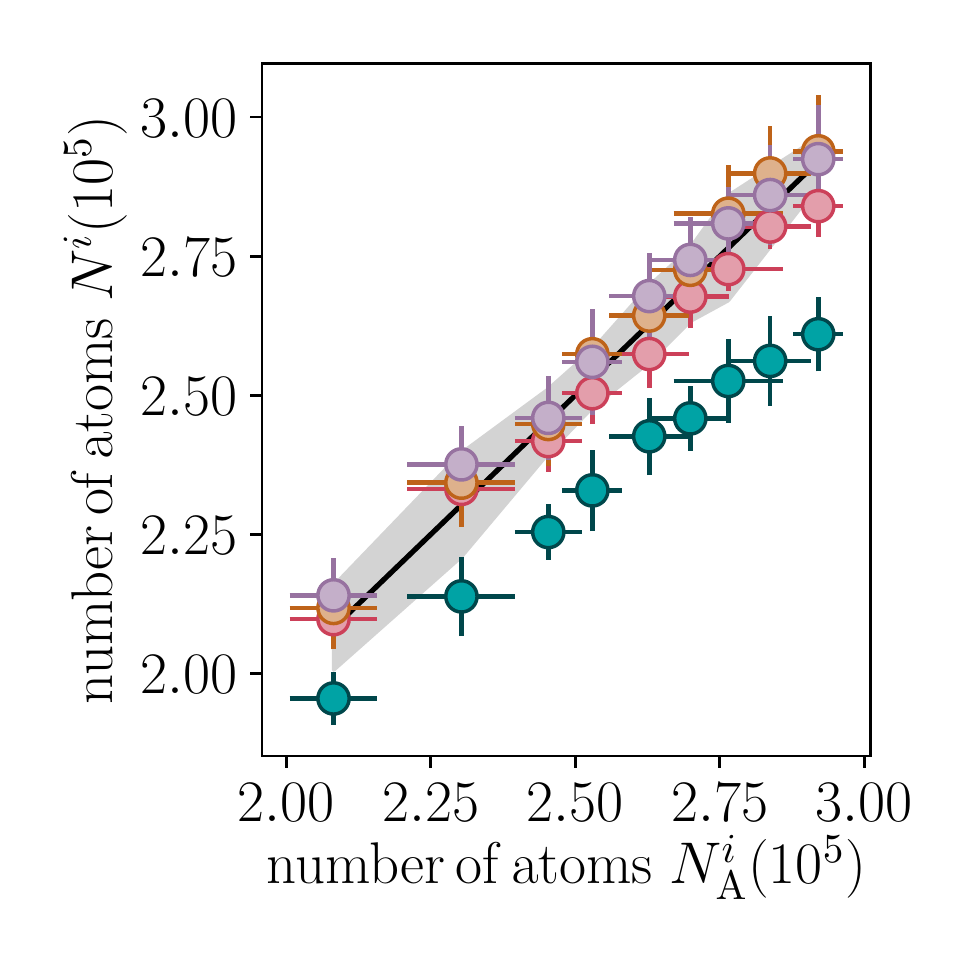}
    \caption{Corrected number of atoms of one spin state $N^i$ in the points of the cycle A (black solid line), B (red), C (orange), D (purple), and $\mathrm{A}_2$ (cyan) in dependence of the corrected number of atoms of one spin state $N^i_{\mathrm{A}}$ in point A. Experimental data belong to the Pauli engine in Fig.~\ref{fig_4_pauli_engine_N}(a) and (c) with a compression ratio of $\bar{\omega}_\mathrm{B}/\bar{\omega}_\mathrm{A}=1.5$. The number of atoms is iterated. The correction factor is already included. Uncertainty is the standard deviation of 20 repetitions.}  
    \label{fig:atom_number}
\end{figure}

Figure \ref{fig:atom_number} shows the measured number of atoms for one spin state $N^i$ for the different points of the cycle of the Pauli engine as a function of the number of atoms of one spin state $N^i_{\mathrm{A}}$ in point A. The correction factors for different magnetic fields are included. 
In this way, we experimentally verify a constant number of atoms during the cycle from A$\rightarrow$D via B and C. Only the last stroke D$\rightarrow$A suffers from atom losses of about \SI{10}{\percent}.

%
%
%
%


%
\begin{figure*}[tb]
    \centering
    \includegraphics[scale=.42]{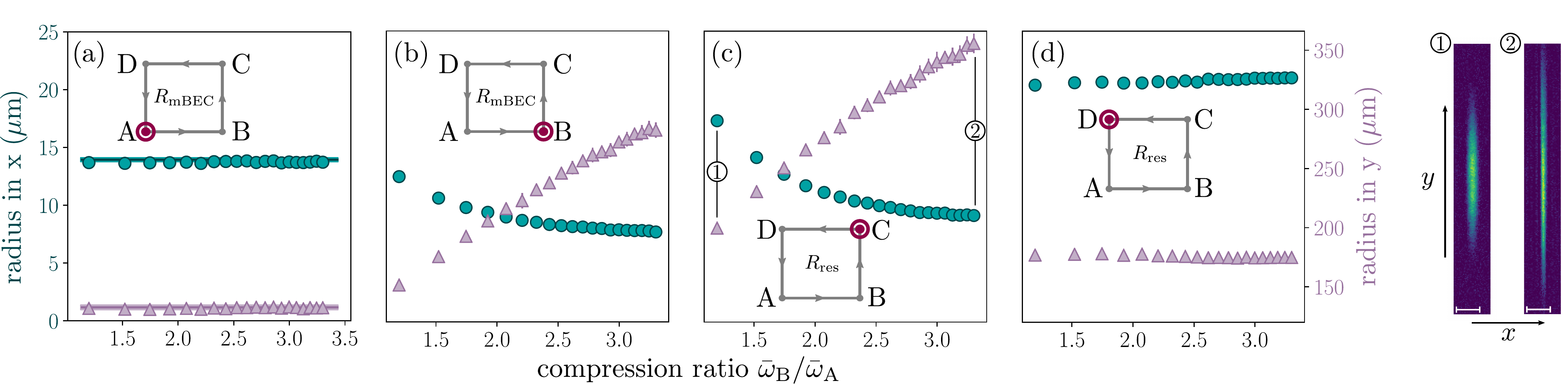}
    \caption{Realization of the Pauli engine. (a)-(d) Measured radii $R_{\mathrm{mBEC}}$ and $R_{\mathrm{res}}$ at each point of the cycle as a function of the  compression ratio $\bar{\omega}_\mathrm{B} / \, \bar{\omega}_\mathrm{A}$ for constant number of atoms $N^i_\mathrm{A}= 2.5 \times 10^5$. Radii are extracted by fits of the density profiles with Eq.~\eqref{eq_TF_fit_mBEC} or ~\eqref{eq_TF_fit_res}. Radii are shown in $x$ direction (cyan circles) and $y$ direction (purple triangles).
    Experimental data are the mean value of 20 repetitions and the corresponding error bars indicate the propagation of the uncertainty of standard deviation of the measurement parameters. The insets show schematics of the cycle, where the point at which the measurement is taken is highlighted (purple).
    (a) Experimental values of the radii for A in the $x$ ($y$) direction are the cyan (purple) line and their uncertainty is the corresponding shaded area (of the size of the line thickness). The trap frequency $\bar{\omega}_\mathrm{B}$ has no influence on the first measurement point A and $\bar{\omega}_\mathrm{A}$ is constant for the different compression ratios. Therefore, this setting is only measured once with 20 repetitions. Experimental data points of radii in A$_2$ (data points) are measured after running a full cycle. On the right two exemplary absorption pictures at point C are shown.}
    \label{fig:appendix_radii}
\end{figure*}
%

\paragraph*{Cloud-radii determination.} 

We obtain line densities of the quantum gases by integrating the column-integrated density distributions of the 2D absorption pictures $n_{\mathrm{2D}}(x,y)=\int n_{\mathrm{3D}}(x,y,z) \, dz$   along the $x$- and $y$-directions separately. We fit these distributions with a 1D-integrated Thomas-Fermi profile. The Thomas-Fermi profiles are different for the interaction regimes throughout the BEC-BCS crossover. In the Thomas-Fermi limit, the in-situ density distribution of an mBEC has the shape \cite{bartenstein_2004}
%
\begin{equation} \label{eq_TF_fit_mBEC}
n_{\mathrm{mBEC}} \propto \left(1-\frac{x^{2}}{R^{2}_{\mathrm{mBEC}_x}}\right)^{2}, 
\end{equation}
%
where $R_{\mathrm{mBEC}_x}$ is the Thomas-Fermi radius of the atomic cloud in $x$ direction. For a resonantly interacting Fermi gas right on resonance, the density profile $n_{\mathrm{res}}$ in the $x$ direction can be written as \cite{giorgini_review_2008}
%
\begin{equation} \label{eq_TF_fit_res}
n_{\mathrm{res}}\propto\left(1-\frac{x^{2}}{R^{2}_{\mathrm{res}_x}}\right)^{5/2}, 
\end{equation}
%
with the rescaled Thomas-Fermi radius $R_{\mathrm{res}_x}$ in the $x$ direction. 
Therefore, we determine the measured cloud radii in both regimes by fitting the measured density profiles with the appropriate line shapes. 
The extracted radii in the $x$ and $y$ directions are shown in Fig.~\ref{fig:appendix_radii}. Our experimental setup does not allow measurements of the radii $R_z$ in the $z$-direction. 
For this radial direction, however, we use the measured radius $R_x$ in the $x$-direction, (radial), and correct the value with the corresponding ratio of the trap frequencies $R_{z} = \frac{\omega_x}{\omega_z} R_x$ \cite{dalfovo_review_1999, BEC_stringari}.


\paragraph*{Temperature measurements.}

We determine the temperature of the mBEC via a bimodal fit of the density profile (for more details, see \cite{doi:10.1063/1.5045827}) at a magnetic field strength of \SI{680}{G}. This value of $B$ is chosen considering the following trade-off: at lower magnetic fields, losses in the atom clouds are too high for quantitative temperature determination, while for higher magnetic fields, the condensate and thermal parts are not well separated, complicating a bimodal fit.  

\begin{figure}[htb]
    \includegraphics[scale=0.6]{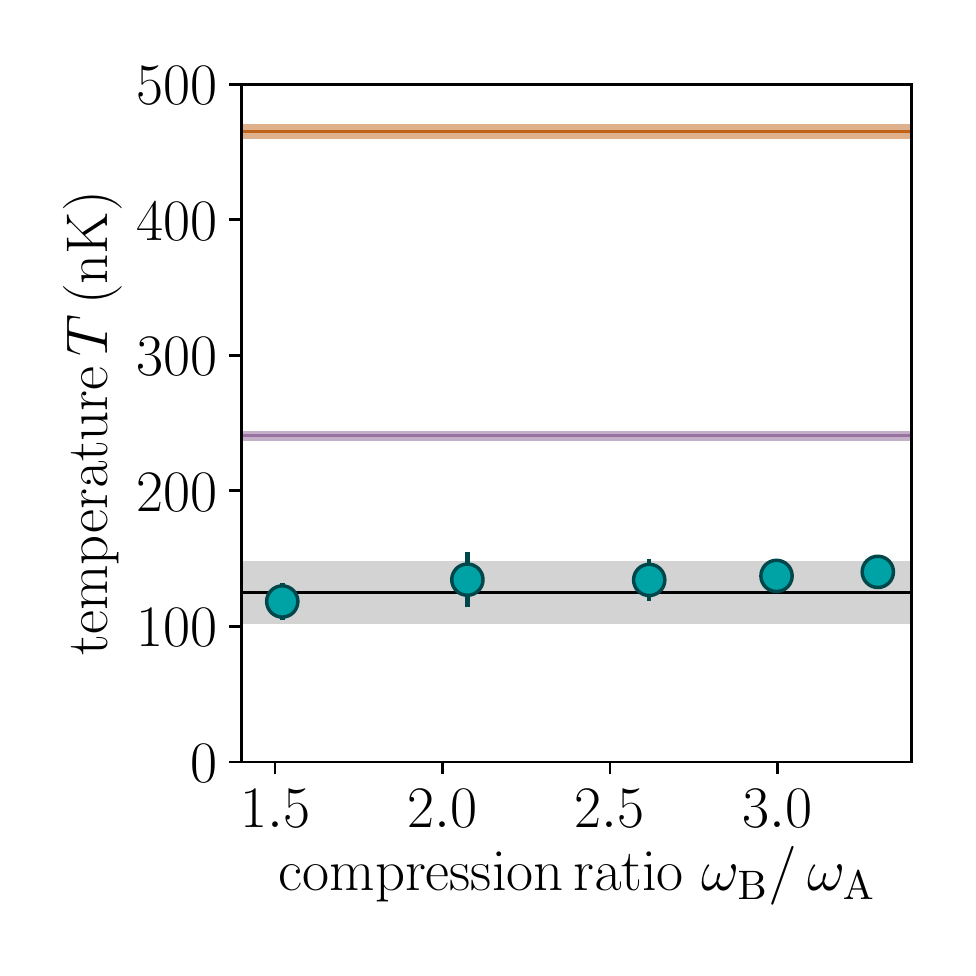}
    \caption{Measured temperature $T$ for point A after ramping the magnetic field directly to \SI{680}{G} (black line). Uncertainty is the standard deviation of ten repetitions (grey area). The cloud is compressed by increasing the mean trap frequency in stroke A$\rightarrow$B. Afterwards, inverting the cycle direction leads again to point A. There, the temperature measurement is repeated after ramping the magnetic field to \SI{680}{G} (cyan data points). Error bars show the standard deviation of ten repetitions. For comparison, critical temperature (purple solid line) and Fermi temperature (orange solid line) are depicted for point A. Shaded area shows the uncertainty and is of the order of the line thickness.}  
    \label{fig:rampT}
\end{figure}

Due to three-body recombination losses in the range between \SI{550}{G} until \SI{750}{G} \cite{phdthesis_jochen}, molecule losses in the cloud are already very high at \SI{680}{G}. To avoid  these losses, we choose a field of \SI{763.6}{G} as start of the Pauli cycle. To determine the temperature in an independent measurement, we ramp the field to 680 G during \SI{200}{ms} and determine the temperature there. However, since decreasing the magnetic field value adiabatically throughout the BEC-BCS crossover increases the temperature $T$ of the gas \cite{williams_njp_2004,chen_thermo_2005}, the measured temperature at \SI{680}{G} is an upper bound for the temperature at point A and beyond.
For the work stroke between points A and B, we increase the mean trap frequency. We observe that the reduced temperature $T/T_{\mathrm{F}}$ of the gas does not show significant changes during these work strokes. The temperature $T$ does not change after ramping the trapping frequency back and forth. 
Due to this, we determine the temperature of the cloud, as mentioned above, with a bimodal fit for the two settings. First, we ramp the magnetic field from point A at \SI{763.6}{G} to \SI{680}{G} to determine the temperature there. 
Second, we restart the sequence and run the cycle until point B. 
Afterwards, we change the direction of the cycle and go directly back to point A. 
A magnetic field sweep to \SI{680}{G} allows again a temperature measurement. 
Figure \ref{fig:rampT} shows that the temperatures of the cloud after reversal to point A (cyan points) lie within the error (grey shaded area) of the initial temperature in point A (black line). 
We observe that an increase of temperature of less than \SI{10}{\percent} takes place for higher ratios of the mean trapping frequency, and we neglect this small increase in the further analysis.

The method described above holds for atomic clouds below the critical temperature. For a thermal gas, we can directly determine the temperature at the magnetic field of the cycle point. At finite temperatures, density profiles of thermal clouds can be approximated with a classical Boltzmann distribution. The temperature $T_j$ can be calculated in dependence of direction $j$ of the cloud $T_j = (m \bar{\omega}^2\sigma_j²)/k_B$ \cite{dalfovo_review_1999}, with $\sigma_j$ as width of a Gaussian determined via fitting the 1D density profiles with $n_{\mathrm{thermal}} \propto \exp(-\frac{j^2}{2\sigma_j^2})$. In the case of molecular Bosons, we use for the mass $2m$ instead of $m$. Hence, interactions directly influence the width of the cloud; we interpret this temperature for our experiment as an approximated temperature.

%
%


\paragraph*{Energy calculation.} 

The calculation of the total energy $E_{\mathrm{mBEC}}$ of a mBEC is based on the Gross-Pitaevskii equation in the Thomas-Fermi limit for a harmonic trapping potential, and for zero temperature. This energy consists of two parts, the kinetic energy $E_{\mathrm{kin}}$, and the energy in the Thomas-Fermi limit $E_{\mathrm{TF}}$, which takes into account the oscillator and interaction energies \cite{dalfovo_review_1999,BEC_stringari}
%
\begin{eqnarray} \label{eq_E_BEC}
    E_{\mathrm{mBEC}} &=& E_{\mathrm{TF}} + E_{\mathrm{kin}} \\
    &=& \left(\frac{\mu_{\mathrm{mBEC}}}{7} + \frac{\hbar^2}{4m R_{\mathrm{mBEC}}^{2}}\left[\text{ln}{\frac{R_{\mathrm{mBEC}}}{1.3a_{\mathrm{ho}}}} + \frac{1}{4}\right]\right)\frac{5N}{2}, \nonumber
\end{eqnarray} 
%
with $R_{\mathrm{mBEC}}$ being the geometric mean radius of the cloud. The chemical potential is given by 
%
\begin{equation}
    \label{eq_mu_BEC}
	\mu_{\mathrm{mBEC}} = \frac{\hbar \bar{\omega}}{2} \biggl( \frac{15\frac{N}{2}a_{\mathrm{dd}}}{a_{\mathrm{ho}}} \biggr)^{2/5},
\end{equation} 
%
where $a_{\mathrm{dd}} = 0.6a$ \cite{Petrov_2004} is the $s$-wave scattering length for molecules, and $a_{\mathrm{ho}} = \sqrt{\hbar/(2m\bar{\omega})}$ the oscillator length. The radii and molecule numbers are extracted from the absorption pictures for the considered interaction strengths.
%
\begin{figure*}[tbh]
    \centering
    \includegraphics[scale=.45]{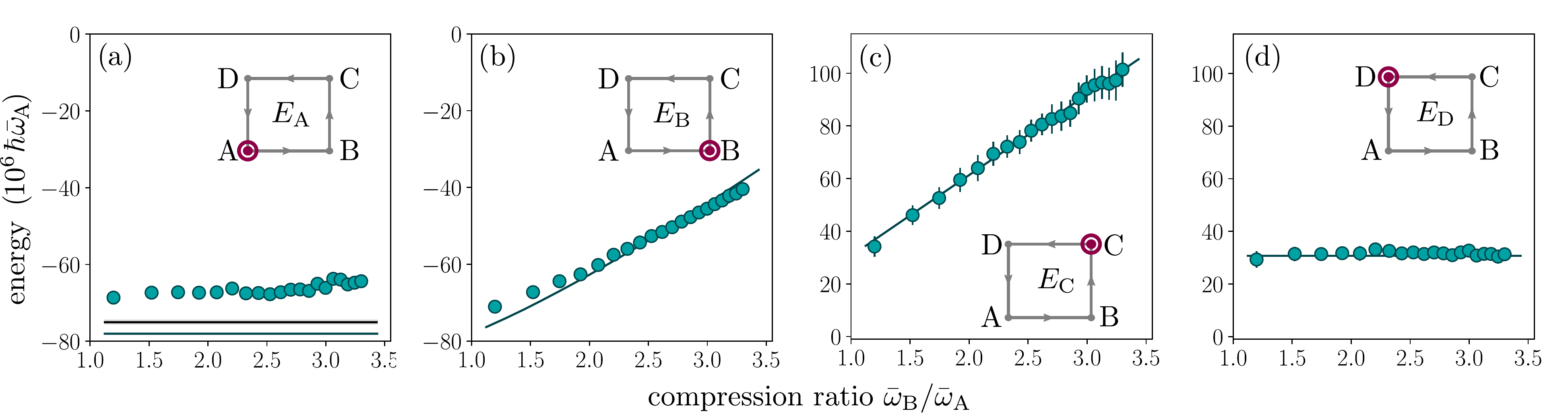}
    \caption{Realization of the Pauli engine. (a)-(d) Energies $E_\mathrm{A}$ and $E_\mathrm{A_2}$, $E_\mathrm{B}$, $E_\mathrm{C}$, and $E_\mathrm{D}$ at each point of the cycle in dependence of the compression ratio  $\bar{\omega}_\mathrm{B} / \, \bar{\omega}_\mathrm{A}$ for constant number of atoms $N^i \approx 2.5 \times 10^5$. Experimental data (cyan points) are the mean value of 20 repetitions and the corresponding error bars indicate the propagation of uncertainty of standard deviation of the measurement parameters. The experimental data fits well to numerical calculations (cyan solid line). The insets show schematics of the engine with the individually points of the engine highlighted in purple. (a) Experimental value of the energy $E_\mathrm{A}$ as a black line and uncertainty as the grey shaded area (of the order of the line thickness). The trap frequency $\bar{\omega}_\mathrm{B}$ has no influence in the first measurement at point A, and $\bar{\omega}_\mathrm{A}$ is constant for the different compression ratios. Therefore, this setting is only measured once with 20 repetitions. Experimental data points of energy $E_\mathrm{A_2}$ (cyan data points) are measured after running a full cycle. }
    \label{fig:appendix_energies}
\end{figure*}
%
When including the molecular energy, the total energy of the mBEC $E_{\mathrm{mBEC,m}}$ is given by
%
\begin{equation}
    \label{eq_app_EBEC}
    E_{\mathrm{mBEC,m}} = E_{\mathrm{mBEC}} - \frac{N}{2} \frac{\hbar^2}{m a^2}. 
\end{equation}
%
On resonance, the scattering length diverges and the binding energy goes to zero. As explained in the main text, we determine the total energy $E_{\mathrm{res}}$ of the trapped cloud at resonance through a scaling with the universal constant $\sqrt{1+\beta}$ 
%
\begin{equation} 
\label{eq:E_unitarity_T0}
	E_{\mathrm{res}}=\frac{3}{4}\sqrt{1+\beta}E_FN,
\end{equation}
%
where $E_F = (3 N)^{1/3}\hbar \bar{\omega}$ is the Fermi energy \cite{Zwerger_book}. This zero-temperature expression provides a good approximation for $T/T_F<0.2$  as supported by the energy measurements on resonance reported in Ref.~\cite{kinast_science_2005}. It should be noted that due to the different temperature dependency of the entropy of the bosonic ($S^{\text{Bosons}}\propto T^3$) and fermionic ($S^{\text{Fermions}}\propto T$) trapped gases, a drastic decrease of the temperature is expected when performing an adiabatic sweep from the mBEC-side to the BCS-side and to unitarity. Since the condition $T/T_F<0.2$ is fulfilled for points C and D in all of our experiments the zero-T formulas on resonance are highly accurate.

The obtained experimental values were contrasted with those obtained using well-known formulas for BECs \cite{dalfovo_review_1999,BEC_stringari} in the Thomas-Fermi limit for interacting bosons adapted as explained before for composite bosons made up of fermions of opposite spin states \cite{greiner_2003, Zwierlein_2003, Jochim_2003}. In these cases we use the theoretically expected radius and a constant number of particles during the cycle equal to the experimental number of particles at point A. Using Eqs.~\eqref{eq_mu_BEC}, \eqref{eq_E_BEC} and $R_{\text{mBEC}}= (9 \hbar^2 N a/(8 m^2 \bar{\omega}^2))^{1/5} $ in Eq.~\eqref{eq_app_EBEC}, we obtain the following energy for the mBEC regime at zero temperature 
\begin{widetext}
\begin{equation}
\label{eq_energy_BEC_zeroT_atomic}
E_{\text{mBEC,m}} = \frac{N}{2} \left\lbrace \frac{5}{14} \hbar \bar{\omega} \left( \frac{9 \frac{N}{2} a}{\sqrt{\frac{\hbar}{2 m \bar{\omega}}}} \right)^{\frac{2}{5}} + \frac{5}{4} \frac{\hbar^2}{m \left( \left( \left( \frac{\hbar^2}{2m \bar{\omega}} \right)^2 9 \frac{N}{2} a \right)^{\frac{1}{5}} \right)^2 } \left[ \text{ln} \left( \frac{10}{13} \left( \frac{9 \frac{N}{2} a}{\sqrt{\frac{\hbar}{2 m \bar{\omega}}}} \right)^{\frac{1}{5}} \right) + \frac{1}{4}  \right]  \right\rbrace - \frac{N}{2} \frac{\hbar^2}{m a^2},
\end{equation} 
\end{widetext}
where the first two terms are related to the energy of bosonic particles in a trap including the interaction between bosons, and the last one is the contribution of the molecular energy of each one of the pairs. For the zero-$T$ calculations, we also numerically solve the Gross-Pitaevskii equation with the bosonic interaction in terms of the dimer-dimer scattering length as explained in the main text. Due to the range of the experimental parameters the Thomas-Fermi approximation holds in all our experiments in the mBEC regime, therefore, the numerical results are the same as the ones obtained when using Eq.~\eqref{eq_energy_BEC_zeroT_atomic}. Figure~\ref{fig:appendix_energies} shows the experimental and theoretical energies for each point of the Pauli cycle for the data set of Fig.~\ref{fig_3_pauli_engine_omega}.

For low but non-zero temperature, the Thomas-Fermi approximation to the Gross-Pitaevskii equation reads \cite{dalfovo_review_1999}
\begin{widetext}
\begin{eqnarray}
\label{eq_energy_BEC_lowT_atomic}
E_{\text{mBEC,m}}^T = \frac{N}{2} \hbar \bar{\omega} \left( \frac{\frac{N}{2}}{\zeta(3)} \right)^{\frac{1}{3}} 
& & \left\lbrace \frac{3 \zeta(4)}{\zeta(3)} \left( \frac{T}{\frac{\hbar \bar{\omega}}{k_B} \left( \frac{\frac{N}{2}}{\zeta(3)} \right)^{\frac{1}{3}}} \right)^4 +\frac{ \zeta(3)^{\frac{1}{3}}}{14} \left( 9 \left( \frac{N}{2}\right)^{\frac{1}{6}} \frac{ a}{\sqrt{\frac{\hbar}{2 m \bar{\omega}}}} \right)^{\frac{2}{5}} \right. 
\nonumber\\
& & \left. \left( 1- \left( \frac{T}{\frac{\hbar \bar{\omega}}{k_B} \left( \frac{\frac{N}{2}}{\zeta(3)} \right)^{\frac{1}{3}}} \right)^3 \right)^{\frac{2}{5}} \left( 5+ 16 \left( \frac{T}{\frac{\hbar \bar{\omega}}{k_B} \left( \frac{\frac{N}{2}}{\zeta(3)} \right)^{\frac{1}{3}}} \right)^3 \right) \right\rbrace - \frac{N}{2} \frac{\hbar^2}{m a^2}.
\end{eqnarray} 
\end{widetext}

According to Tan's generalized virial theorem \cite{TAN20082987, Liu_2009,Hu_2010,Hu_2011,Zwerger_book}, the energy for a cigar-shaped trapped Fermi gas is given by $E = N m \left( 2 \omega_r^2 \left\langle r \right\rangle ^2+ \omega_z^2 \left\langle z \right\rangle ^2 \right) - \hbar^2 {\cal I}/(8 \pi m a)$ where $r$ and $z$ stand for the radial and axial coordinates respectively and ${\cal I}$ is the contact (a measure of the probability for two fermions of opposite spins of being close together \cite{Zwerger_book}). This expression is valid for any value of the scattering length $a$ and for any temperature $T$. At unitarity we have $1/k_F a = 0$ and a finite ${\cal I}$, therefore the virial theorem reduces to the case presented by Thomas \textit{et al.} in Refs.~\cite{Thomas_2005,PhysRevA.78.013630}, i.e. $E = N m \left( 2 \omega_r^2 \left\langle r \right\rangle ^2+ \omega_z^2 \left\langle z \right\rangle ^2 \right)$. In the weak coupling limit for the mBEC-side the contact reduces to ${\cal I} \sim 4\pi N/a$, case in which the last term of the virial expression gives the total molecular energy and the virial energy reduces to Eqs.~\eqref{eq_energy_BEC_zeroT_atomic} and \eqref{eq_energy_BEC_lowT_atomic} \cite{TAN20082987, dalfovo_review_1999}. All of this means that the expression $- \hbar^2/(m a^2)$ for the molecular term holds even for $T/T_F \approx 0.7$ as long as the experiments are realized in the deep mBEC regime. For our magnetic fields this condition is fulfilled for the Pauli and mBEC-mBEC engine as well as for the thermal cycle. Furthermore, measurements of the molecular energy were presented in Ref.~\cite{regal_2003_nature} for $T/T_F \approx 0.15$ showing that the formula $- \hbar^2/(m a^2)$ holds for $T/T_F \approx 0.2$ in a complete $B$ sweep from mBEC to unitarity.

In the high-$T$ regime the interactions are negligible and the distributions can be approximated by Maxwell-Boltzmann distributions. When the thermal energy $k_B T$ is large enough to break all the pairs the energy can be approximated by that of an ideal classical gas of atomic particles for both magnetic fields. In this case the temperature in the work strokes does not change and therefore there is no work output. When the thermal energy is below the binding energy, the energy for a magnetic field below resonance corresponds to a classical gas of molecules with mass $2m$ while the energy at the resonant field is given by a classical gas of atoms in a harmonic trap. The molecular term cancels in the total work output giving $W = 3 k_B N \left( T_\text{A}/2 - T_\text{D} + T_\text{C} - T_\text{B}/2 \right)$. The temperature of the gas is obtained through a Gaussian fitting of the density profile leading to $k_B T = m \bar{\omega}^2 \sigma^2$ with $\sigma$ being the width of the fitted density profile. Since $\bar{\omega}_\text{D} = \bar{\omega}_\text{A}$ and $\bar{\omega}_\text{C} = \bar{\omega}_\text{B}$ and the mass of the molecules is twice the one of the atoms, we have that $T_\text{A}/2 = (\sigma_\text{A}/\sigma_\text{D})^2 T_\text{D}$ and $T_\text{B}/2 = (\sigma_\text{B}/\sigma_\text{C})^2 T_\text{C}$, i.e. $W = 3 k_B N \left\lbrace \left( \sigma_\text{A}^2/\sigma_\text{D}^2 - 1 \right) T_\text{D} + \left( 1 - \sigma_\text{B}^2/\sigma_\text{C}^2 \right) T_\text{C} \right\rbrace$. The work vanishes for $\sigma_\text{A} = \sigma_\text{D}$ and $\sigma_\text{B} = \sigma_\text{C}$. In our experiments the widths do not show significant statistical differences and therefore the work output for the engine running in the high-$T$ regime vanishes.

Let us finally detail each of the theoretical curves presented in the main text. Based on Tan's generalized virial theorem we calculate the theoretical potential energy of Fig.~\ref{fig_2_pauli_stroke} via Eq.~\eqref{eq_energy_BEC_lowT_atomic} without the molecular term and using the experimental temperatures at points A and B, while for the points C and D we use the energy for zero-$T$ at unitarity given by Eq.~\eqref{eq:E_unitarity_T0}. The vanishing efficiency for the thermal case is expected from the classical gas argument of the previous paragraph. The cyan curves in Fig.~\ref{fig_3_pauli_engine_omega} arise from numerical calculations of the Gross-Pitaevskii equation. The black dashed curves in Fig.~\ref{fig_3_pauli_engine_omega} (e)-(f) were also calculated by solving numerically the Gross-Pitaevskii equation and must give the same results as an ideal molecular gas of point-like bosons having therefore an infinite binding energy, which leads to $W_\text{non-interacting} = (3 N)^{4/3} \sqrt{1+\beta} \hbar (\bar{\omega}{_\text{B}}-\bar{\omega}{_\text{A}})/4$ and $\eta_\text{non-interacting} = 0$. The purple curves in Figs.~\ref{fig_3_pauli_engine_omega}(e) and (f) correspond to the results obtained for ideal Bose and Fermi gases in the highly degenerate regime and lead to $W_\text{ideal} = (3 N)^{4/3} \hbar (\bar{\omega}{_\text{B}}-\bar{\omega}{_\text{A}})/4$ and $\eta_\text{ideal} = 1 -\bar{\omega}{_\text{B}}/\bar{\omega}_{\text{A}}$. The gray curves in Figs.~\ref{fig_3_pauli_engine_omega} (e) and (f) and the theoretical curves of Fig.~\ref{fig_4_pauli_engine_N} follow from Eqs.~\eqref{eq_energy_BEC_zeroT_atomic} and \eqref{eq:E_unitarity_T0}. As mentioned in the main text, since the latter calculations rely on zero-$T$ formulas we interpret their results as an upper bound for the work output and efficiency. To show that this estimation is accurate we need to consider two aspects. First, for the two on-resonance points (C,D) we are certain that the zero-$T$ calculations constitute a good approximation because of the aforementioned drastic decay in the temperature between the mBEC regime and unitarity. Due to the dominance of the molecular term on the mBEC-side the efficiency obtained when including the temperature overlap with the one obtained within the zero-$T$ approximation. 
For the data of Fig.~\ref{fig_4_pauli_engine_N} the zero-$T$ calculations give an efficiency of about 0.07(2) while the finite-$T$ formulas give an efficiency of about 0.05(3). A null hypothesis test for the mean difference gives a $p-$value of $0.1$. Setting the usual significance of $\alpha=0.05$ the null hypothesis of equal efficiencies can not be rejected. We therefore conclude that the finite-T corrections have no significant effects on the efficiency of our engine.


\bibliography{bib_pauli_engine}
 

\end{document}